\documentclass[doublecol]{epl2}

\usepackage{amssymb}
\usepackage{amsmath}
\usepackage{graphicx}
\usepackage{subfigure}

\title{Fluctuations of the superconducting order parameter as an origin of the Nernst effect}
\shorttitle{Superconducting fluctuations as an origin of the Nernst effect}

\author{K. Michaeli\inst{1} \and A. M. Finkel'stein\inst{1,2}}

\institute{
  \inst{1} Department of Condensed Matter Physics, The Weizmann Institute of Science,
Rehovot 76100, Israel \\
  \inst{2} Department of Physics, Texas A\&M University, College Station, TX $77843-4242$, USA
}

\abstract{We show that the strong Nernst signal observed recently in amorphous superconducting films far above the critical temperature is caused by the fluctuations of the superconducting order parameter. We demonstrate a striking agreement between our theoretical calculations and the experimental data at various temperatures and magnetic fields.  Besides, the Nernst effect is interesting not in the context of superconductivity only. We discuss some subtle issues in the theoretical study of thermal phenomena that we have encountered while calculating the Nernst coefficient. In particular, we explain how the Nernst theorem (the third law of thermodynamics) imposes a strict constraint on the magnitude of the Nernst effect.}

\begin{document}

\maketitle

The recent observations of a transverse thermoelectric signal in the presence of a magnetic field (the Nernst effect) well above the critical temperature of the superconducting transition ($T_c$) both in conventional and high-$T_c$ material have made this phenomenon an area of high theoretical interest~\cite{Zaanen2007,Sachdev2007}. The Nernst effect in high-$T_c$ superconductors~\cite{Ong2000,Ong2005} has been attributed to the motion of vortices~\cite{Huse2004,Podolsky2007,Anderson2007} existing even above $T_c$ (the vortex-liquid regime). In conventional amorphous superconducting films the strong Nernst signal observed deep in the normal state~\cite{Aubin2006,Aubin2007} cannot be explained by the vortex-like fluctuations. Rather, the authors of Refs.~\cite{Aubin2006,Aubin2007} suggested that the effect is caused by fluctuations of the superconducting order parameter. In this letter, we analyze this mechanism and demonstrate a quantitative agreement between the theoretical expressions and the experiment~\cite{Aubin2007}. No fitting parameters have been used; the values of $T_c$ and the diffusion coefficient have been taken from independent measurements (see Refs.~\cite{Aubin2006,Aubin2007}).
In particular, we succeeded in reproducing the non-trivial dependence of the signal on the magnetic
field. Our results imply that in the quest for understanding the thermoelectric phenomena in high-$T_c$ materials
the fluctuations of the order parameter should also not be ignored.

In metallic conductors the quasi-particle excitations yield a negligible contribution to the Nernst effect and to its counterpart, the Ettingshausen effect. Under the approximation of a constant density of states at the Fermi energy, which is a standard approximation for the Fermi liquid theory, this contribution vanishes completely
~\cite{Sondheimer1948}. On the other hand, there is no general reason why the collective modes describing all kinds of fluctuations should not contribute to the Nernst effect (in our opinion, the opposite statement in Ref.~\cite{Reizer2008} has not been justified). Obviously, neutral modes are not deflected by the Lorentz force and cannot contribute to the transverse-voltage signal. However, the charged modes, such as the fluctuations of superconducting order parameter, can be a source for the strong Nernst effect even far from the superconducting transition.

\begin{figure}[pt]
        \onefigure[width=0.3\textwidth]{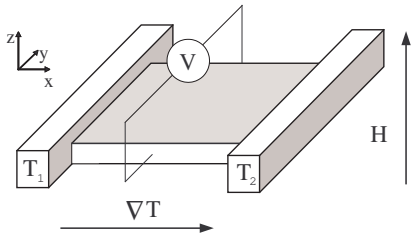}
                 \caption[0.4\textwidth]{\small The setup of the Nernst effect measurement. The sample is placed between two thermal baths of different temperatures. The temperature gradient is in the $x$-direction, the magnetic field is along the $z$-direction and the electric field is induced in the $y$-direction.} \label{fig:NernstSetup}
\end{figure}

The contributions to the electric conductivity caused by the superconducting fluctuations (paraconductivity) have already been known for $40$ years~\cite{Aslamazov1968,Maki1968,Varlamov}. Close enough to the superconducting transition the paraconductivity increases rapidly and may even overcome the Drude conductivity. Far from the transition the superconducting fluctuations produce only one among many corrections to the conductivity and, therefore, can hardly be identified. Owing to the fact that in the absence of fluctuations the Nernst effect is negligible, measurements of the Nernst signal provide a unique opportunity to study the superconducting fluctuations deep inside the normal state.

The transport coefficients for the electric and heat currents are defined via the standard conductivity tensor:
\begin{equation}
\left(\begin{array}{c}
\mathbf{j}_{e} \\
\mathbf{j}_{h}\end{array}\right)=
\left(
\begin{array}{cc}
\hat{\sigma} & \hat{\alpha} \\
\hat{\tilde{\alpha}} & \hat{\kappa} \\
\end{array}\right)\left(\begin{array}{c}
\mathbf{E} \\
-\boldsymbol{\nabla}{T}
\end{array}\right).
\label{eq:ConductivityTensor}
\end{equation}%
When the thermo-magnetic phenomena are studied in films (or layered conductors) the magnetic field is conventionally directed perpendicularly to the conducting plane, see Fig.~\ref{fig:NernstSetup}. Then  each element of the conductivity tensor corresponds to a $2\times2$ matrix describing the conductivity components in the $x-y$ plane (see Fig.~\ref{fig:NernstSetup}). The Onsager relations imply that $\sigma_{ij}(\mathbf{H})=\sigma_{ji}(-\mathbf{H})$ and $\tilde{\alpha}_{ij}(\mathbf{H})=T{\alpha}_{ji}(-\mathbf{H})$. From the condition, $\mathbf{j}_e=0$, one gets that the Nernst coefficient is
\begin{equation}
e_{N}=\frac{E_{y}}{-\nabla_{x}{T}}=\frac{\sigma_{xx}\alpha_{xy}-\sigma_{xy}\alpha_{xx}}{\sigma_{xx}^{2}+\sigma_{xy}^2}.
\label{eq:NernstCoefficient}
\end{equation}%
We checked that the second term in the numerator is negligible in comparison to the first one (see the comment below Eq.~\ref{eq:Magnetization}). Therefore, the leading order term for the Nernst coefficient is $e_{N}\approx\alpha_{xy}/\sigma_{xx}$ and our goal is to find the transverse Peltier coefficient, $\alpha_{xy}$.

In the linear regime, the electric current generated as a response to an external force, such as the electric field, can be found using the Kubo formula~\cite{Kubo1957} which expresses the response in terms of a correlation function. Extending the Kubo formalism to the calculation of the response to the temperature gradient is not trivial because this gradient is not directly connected to any mechanical force. Following the scheme used in the derivation of the Einstein relation, Luttinger~\cite{Luttinger1964} made a connection between the responses to the temperature gradient and to an artificial gravitational field. As a result, Luttinger succeeded to relate all transport coefficients with various current-current correlation functions. However, in the presence of interactions, the expression for the heat current becomes a non-trivial function of the interaction. Usually, instead of the full expression for the heat current operator, the operator of a non-interacting electrons is used (see e.g., Ref.~\cite{Smith}). Unfortunately, such a simplified version of the Kubo formula has no justification for the problem considered here. In the presence of a magnetic field, the Kubo formalism meets with an additional difficulty when the thermoelectric currents are considered. Obraztsov~\cite{Obraztsov1965} has pointed out that  the heat current describing the flow of entropy must also include a contribution from the magnetization~\cite{Obraztsov1965,Streda1977,Halperin1997}. Thus, the problem of this approach is that the current cannot be expressed by a correlation function alone.

In the derivation of the thermoelectric currents we have decided, instead of applying the Kubo formula, to employ a different approach and to use the quantum kinetic equation~\cite{Keldysh1964,Rammer1986,Haug}. One main advantage of this approach as we will show is that the problem of the magnetization current has been automatically solved. In this way, we obtain that the Peltier and Nernst coefficients, which are related to the flow of entropy, vanish as $T\rightarrow0$ in accordance with the third law of thermodynamics~\cite{Hu1976}. As we will see, the third law of thermodynamics imposes a strict constraint on the structure of the different contributions to the Peltier coefficient.~\footnote{Besides the Peltier coefficient, we may independently find the heat current induced by an applied electric field. Thus, the  quantum kinetic approach allows us to calculate separately both off-diagonal components of the conductivity tensor  (see Eq.~\ref{eq:ConductivityTensor}) and directly  verify the Onsager relations.}

\section{The quantum kinetic approach}

In the presence of superconducting fluctuations we describe the system using two fields. One is the quasi-particles field  $\psi$, while the other represents the fluctuations of the superconducting order parameter $\Delta$. The matrix functions $\hat{G}(\mathbf{r},\mathbf{r}',\epsilon)$ and $\hat{L}(\mathbf{r},\mathbf{r}',\omega)$ written in the Keldysh form~\cite{Keldysh1964,Rammer1986,Haug} describe the propagation of these two fields, respectively. The propagators depend on the two spatial coordinates because we postpone the averaging over the disorder until the last stage of the calculation. The $\boldsymbol{\nabla}T$-dependent part of the quasi-particles Green function is naturally separated into two terms. The first one describes the readjustment of quasi-particles to the non-uniform temperature when the system is trying to maintain a local equilibrium. In the regime of linear response, this local-equilibrium Green function becomes:
\begin{align}\label{eq:G_LocEq}
&\hat{G}_{loc-eq}(\mathbf{r},\mathbf{r}',\epsilon)=-\frac{(\mathbf{r+r}')\boldsymbol{\nabla}T}{2T}\epsilon\frac{\partial\hat{g}(\mathbf{r},\mathbf{r}',\epsilon)}{\partial\epsilon}.
\end{align}
Here $\hat{g}(\mathbf{r},\mathbf{r}',\epsilon)$ is the equilibrium quasi-particles Green function at a constant temperature $T$. We see that the local equilibrium Green function is a straightforward extension of the equilibrium Green function for a non-uniform temperature. Since the same relation holds for the equilibrium and local equilibrium self-energies, $\hat{\sigma}$ and $\hat{\Sigma}_{loc-eq}$, the equation for $\hat{G}_{loc-eq}$ is a closed equation fully determined by the equilibrium properties of the system.

The other term in the $\boldsymbol{\nabla}T$-dependent part of the Green function is
\begin{align}\label{eq:G_TransInv}
\hat{G}_{\boldsymbol{\nabla}T}&(\mathbf{r},\mathbf{r}',\epsilon)=\hat{g}\left(\epsilon\right)
\hat{\Sigma}_{\boldsymbol{\nabla}T}\left(\epsilon\right)\hat{g}\left(\epsilon\right)\\\nonumber
&-\frac{i\boldsymbol{\nabla}T}{2T}\epsilon\left[\frac{\partial\hat{g}\left(\epsilon\right)}{\partial\epsilon}\mathbf{\hat{v}}(\epsilon)\hat{g}\left(\epsilon\right)
-\hat{g}\left(\epsilon\right)\mathbf{\hat{v}}(\epsilon)\frac{\partial\hat{g}\left(\epsilon\right)}{\partial\epsilon}
\right].
\end{align}
The product of matrices should be understood as a convolution in real space. The matrix $\mathbf{\hat{v}}(\mathbf{r},\mathbf{r}',\epsilon)$ is the velocity of the quasi-particles at equilibrium renormalized by the self-energy $\hat{\sigma}(\mathbf{r},\mathbf{r}',\epsilon)$:
\begin{align}\label{eq:VelocityQP}
\mathbf{\hat{v}}(\mathbf{r},\mathbf{r}',\epsilon)&=-\frac{i}{2m}\lim_{\mathbf{r}'\rightarrow\mathbf{r}}\left(\boldsymbol{\nabla}-\frac{ie}{c}\mathbf{A}(\mathbf{r})-\boldsymbol{\nabla}'-\frac{ie}{c}\mathbf{A}(\mathbf{r}')\right)\\\nonumber
&
-i(\mathbf{r-r}')\hat{\sigma}(\mathbf{r},\mathbf{r}',\epsilon),
\end{align}
where $\mathbf{A}(\mathbf{r})$ is the vector potential.

Let us point out an important difference between the two parts of the Green function depending on the temperature gradient, $\hat{G}_{loc-eq}$ and $\hat{G}_{\boldsymbol{\nabla}T}$. As has been already mentioned, $\hat{G}_{loc-eq}$ and $\hat{\Sigma}_{loc-eq}$ are a straightforward extension of the equilibrium Green function and self-energy for a non-uniform temperature. On the other hand, the equation for $\hat{G}_{\boldsymbol{\nabla}T}$ contains the self-energy $\Sigma_{\boldsymbol{\nabla}T}$ which by itself is a function of $\hat{G}_{\boldsymbol{\nabla}T}$. Thus, this is a self consistent equation, and in order to find a close expression for $\hat{G}_{\boldsymbol{\nabla}T}$, one has to determine the structure of the self-energy.

The equations for the $\boldsymbol{\nabla}T$-dependent parts of the superconducting fluctuations propagator remind the first term in Eq.~\ref{eq:G_TransInv} for $\hat{G}_{\boldsymbol{\nabla}T}$:
\begin{align}\label{eq:L_loceq}
&\hat{L}_{loc-eq}(\mathbf{r},\mathbf{r}',\omega)=-\hat{L}(\omega)\hat{\Pi}_{loc-eq}(\omega)\hat{L}(\omega);\\\nonumber
&\hat{L}_{\boldsymbol{\nabla}T}(\mathbf{r},\mathbf{r}',\omega)=-\hat{L}(\omega)\hat{\Pi}_{\boldsymbol{\nabla}T}(\omega)\hat{L}(\omega).
\end{align}
Here $\hat{L}$ is the equilibrium propagator of the fluctuations at temperature $T$. The self-energy of the superconducting fluctuations $\Pi(\mathbf{r},\mathbf{r}',\omega)$ depends on the temperature gradient through the quasi-particle Green functions. Similar to Eq.~\ref{eq:VelocityQP}, we define $\hat{\boldsymbol{\mathcal{V}}}(\mathbf{r},\mathbf{r}',\omega)$ to be the "renormalized velocity" of the superconducting fluctuations (with a contact s-wave interaction):
\begin{equation}\label{eq:VelocitySC}
\boldsymbol{\hat{\mathcal{V}}}(\mathbf{r},\mathbf{r}',\omega)=-i(\mathbf{r-r'})\hat{\Pi}(\mathbf{r},\mathbf{r}',\omega).
\end{equation}
Note that in fact $\hat{\boldsymbol{\mathcal{V}}}$ does not have the dimension of velocity.

For the calculation of the Nernst effect, we need to obtain the expression for the electric current in terms of the $\boldsymbol{\nabla}T$-dependent propagators. In the presence of a magnetic field, the electric current is a sum of two terms:
\begin{align}\label{eq:Jtr+Jmag}
\mathbf{j}_e&=\mathbf{j}_{e}^{con}+\mathbf{j}_{e}^{mag}.
\end{align}
The current $\mathbf{j}_{e}^{con}$ can be as usually derived through the continuity equation for the electric charge. Since the field $\Delta$ carries charge, it is obvious that the charge of the quasi-particles, $-e|\psi(\mathbf{r})|^2$, is not conserved unless the current carried by the superconducting fluctuations is also included. Then, we obtain using the expressions for the $\boldsymbol{\nabla}T$-dependent propagators that
\begin{widetext}
\begin{align}\label{eq:JETransInv}
j_{e\hspace{1.5mm}i}^{con}&=
-\frac{e\nabla_{j}T}{2T}\int\frac{d\epsilon}{2\pi}\epsilon\frac{\partial{n_F(\epsilon)}}{\partial\epsilon}\left[v_{i}^R(\epsilon)g^{R}(\epsilon)v_{j}^A(\epsilon)g^A(\epsilon)
+v_{i}^R(\epsilon)g^{R}(\epsilon)v_{j}^R(\epsilon)g^A(\epsilon)-v_{i}^R(\epsilon)g^{R}(\epsilon)v_{j}^R(\epsilon)g^R(\epsilon)\right.\\\nonumber
&\left.-
g^{R}(\epsilon)v_{j}^R(\epsilon)g^R(\epsilon)v_{i}^A(\epsilon)
\right]-
\frac{e\nabla_jT}{T}\int\frac{d\epsilon}{2\pi}\epsilon{n}_F(\epsilon)\left[v_{i}^R(\epsilon)\frac{\partial{g}^R(\epsilon)}{\partial\epsilon}v_{j}^R(\epsilon)g^R(\epsilon)-
-v_{i}^R(\epsilon){g}^R(\epsilon)v_{j}^R(\epsilon)\frac{\partial{g}^R(\epsilon)}{\partial\epsilon}\right]\\\nonumber
&-{ie}\int\frac{d\epsilon}{2\pi}v_{i}^R(\epsilon)g^R(\epsilon)\left[\Sigma_{\boldsymbol{\nabla}T}^{<}(\epsilon)(1-n_F(\epsilon))+\Sigma_{\boldsymbol{\nabla}T}^{>}(\epsilon)n_F(\epsilon)\right](g^{R}(\epsilon)-g^A(\epsilon))\\\nonumber
&+ie\hspace{-1mm}\int\hspace{-1mm}\frac{d\omega}{2\pi}\mathcal{V}_{i}^R(\omega)L^R(\omega)\left[\Pi_{\boldsymbol{\nabla}T}^{<}(\omega)(1+n_P(\omega))-\Pi_{\boldsymbol{\nabla}T}^{>}(\omega)n_P(\omega)\right](L^{R}(\omega)-L^A(\omega))+c.c.
\end{align}
\end{widetext}
\begin{floatequation}\nonumber
\mbox{\textit{see eq.~\eqref{eq:JETransInv}}}
\end{floatequation}
Note that the local equilibrium propagators do not contribute to $\mathbf{j}_{e}^{con}$.

\begin{figure}[pt]
        \onefigure[width=0.35\textwidth]{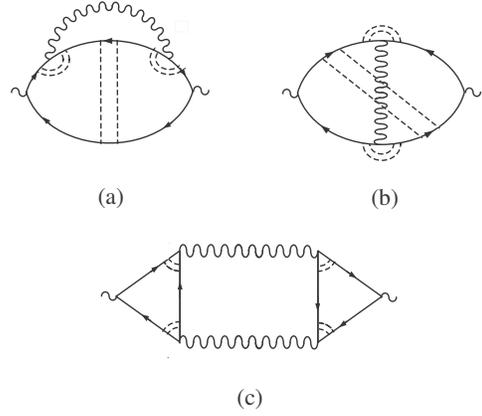}
                 \caption[0.4\textwidth]{\small  The diagrammatic contributions to the thermoelectric current. Diagrams a) and b) describe the fluctuation of the superconducting order parameter decorated by three Cooperons and c) is the Aslamazov-Larkin diagram. (The obvious counterpart diagrams for a) and b) are not shown.) These contributions should be supplemented by the magnetization current term.} \label{fig:diagrams}
\end{figure}

Our only assumption in the derivation of Eq.~\ref{eq:JETransInv} is that the current is obtained in the regime of linear response to $\boldsymbol{\nabla}T$. As we are interested in the Gaussian fluctuations, we expand the expression for the current with respect to the interaction with the superconducting fluctuations. In Fig.~\ref{fig:diagrams}, we give a diagrammatic interpretation for the dominant contributions to the transverse thermoelectric current after averaging over the disorder. The analytical structure and the expression for the vertices of these diagrams have been found from the quantum kinetic equation. In principle, the same diagrams can be calculated using the Kubo formula. However, if for simplicity one uses in the Kubo formula the heat current operator of non-interacting electrons, the resulting expressions for these diagrams differ from those obtained using the quantum kinetic equation. Most important, as one can see from Eq.~\ref{eq:JETransInv}, in the quantum kinetic approach the frequency accompanies the renormalized velocity, so that the expression for the electric current is generally of the form $eg(\epsilon)v_i(\epsilon)g(\epsilon)\epsilon{v}_j(\epsilon)\nabla_jT/T$. In other words, the frequency appears together with the velocity that has been already renormalized by the interaction. On the other hand, owing to the fact that the frequency in the simplified version of the Kubo formula is attached to the external vertex before the renormalization of the velocity, the expression for the current has a totaly different structure.

The second contribution to the electric current is from the magnetization current. Since the magnetization current is
divergenceless, it cannot be obtained using the continuity equation. Rather, it is found directly from the action:
\begin{align}\label{eq:Jmag}\tag{10}
\mathbf{j}_{e}^{mag}&=-{2ic}\boldsymbol{\nabla}\times\mathbf{M}(\mathbf{r})\\\nonumber
&\lim_{\mathbf{r}'\rightarrow\mathbf{r}}\int\frac{d\epsilon}{2\pi}
\left[\hat{G}_{loc-eq}(\mathbf{r}';\mathbf{r},\epsilon)+\hat{G}_{\boldsymbol{\nabla}T}(\mathbf{r}';\mathbf{r},\epsilon)\right]^{<},
\end{align}
where $\mathbf{M}(\mathbf{r})$  denotes the magnetization and the factor of $2$ is due to the summation over the spin index. It can be checked that $\hat{G}_{\boldsymbol{\nabla}T}$ does not contribute to $\mathbf{j}_{e}^{mag}$. On the other hand, the explicit dependence of the local equilibrium Green function on the center of mass coordinate leads to a non-zero contribution to the magnetization current:
\begin{align}\label{eq:Jmag2}
\mathbf{j}_e^{mag}&=2ic\boldsymbol{\nabla}\times\mathbf{M}(\mathbf{r})\lim_{\mathbf{r}'\rightarrow\mathbf{r}}\int\frac{d\epsilon}{2\pi}\epsilon\frac{(\mathbf{r+r}')\boldsymbol{\nabla}T}{2T}\frac{\partial{g}^{<}(\mathbf{r},\mathbf{r}',\epsilon)}{\partial\epsilon}.
\end{align}
Thus, $G_{\boldsymbol{\nabla}T}$ and $G_{loc-eq}$ are complementary to each other; while the first contributes only to $\mathbf{j}_{e}^{con}$, the other one fully determines $\mathbf{j}_{e}^{mag}$.
One should recall that we are looking for a current that does not vanish after spatial averaging, i.e., after integration with respect to the center of mass coordinate $\mathbf{r}$. Since in the process of averaging over $\mathbf{r}$ we may integrate by parts, the magnetization current can be written as
\begin{align}\label{eq:Jmag3}
\mathbf{j}_{e\hspace{1.5mm}i}^{mag}&=2i\varepsilon_{ij}\frac{\nabla_jT}{T}cM_z\lim_{\mathbf{r}'\rightarrow\mathbf{r}}\int\frac{d\epsilon}{2\pi}{g}^{<}(\mathbf{r},\mathbf{r}',\epsilon)\\\nonumber
&
\equiv-\varepsilon_{ij}c\langle{M_z}\rangle\frac{\nabla_jT}{T},
\end{align}
where $\varepsilon_{ij}$ is the anti-symmetric tensor. In the transition between Eqs.~\ref{eq:Jmag2} and~\ref{eq:Jmag3} we integrated by parts over the frequency as well. The result demonstrates the strength of the quantum kinetic approach. This method provides a way to derive both components of the current quantum mechanically without engaging any thermodynamical arguments. At this point we may employ in Eq.~\ref{eq:Jmag3} the known expression for the magnetization in the presence of superconducting fluctuation:
\begin{align}\label{eq:Magnetization}
&\mathbf{j}_{e\hspace{1.5mm}i}^{mag}=\varepsilon_{ij}\frac{\nabla_jT}{T}\frac{\partial}{\partial{H}}\frac{eH}{\pi}\sum_{N=0}^{\infty}\sum_{\omega_m=-\infty}^{\infty}\ln\left[L_N^{-1}(\omega_m)\right];\\\nonumber
&L_N^{-1}(\omega_m)=-\nu\left[\ln\frac{T}{T_c}+\psi\left(\frac{1}{2}+\frac{|\omega_m|+\Omega_c(N+1/2)}{4\pi{T}}\right)\right.\\\nonumber
&\left.\hspace{50mm}-\psi\left(\frac{1}{2}\right)+i\varsigma\omega_m\right].
\end{align}
Here $\Omega_c=4eHD/c$ is the cyclotron frequency for the collective mode of the fluctuation of the superconducting order parameter where $D$ is the diffusion coefficient. The parameter $\varsigma\propto1/(g\varepsilon_F)$  is important for understanding the difference in magnitude between the longitudinal and transverse Peltier coefficients ($\varepsilon_F$  is the Fermi energy while $g$ is the dimensionless coupling constant determining $T_c$). The longitudinal Peltier coefficient, $\alpha_{xx}$, contains an integral over the frequency that vanishes when $\varsigma=0$ while the integrand that determines $\alpha_{xy}$ remains finite even in the absence of $\varsigma$.  As a result, in the expression for the Nernst coefficient given in  Eq.~\ref{eq:NernstCoefficient}, the second term in the numerator is smaller than the first one by a factor of the order $T/(g\varepsilon_F)$~\cite{Larkin1995}.

Further analysis of $\mathbf{j}_{e}^{con}$ and $\mathbf{j}_{e}^{mag}$ at arbitrary temperatures and magnetic fields shows that they have contributions where the frequency integration accumulates over a wide interval between $T$ and $1/\tau$ (the scattering rate of electrons by impurities). The outcome of the integration depends logarithmically on $1/\tau$ that acts as an ultraviolet cutoff. In addition, as the temperature goes to zero, there is even a more serious problem with these terms because their pre-factor is proportional to $\Omega_c/T$.  We have found a way to verify that the logarithmic parts of $\mathbf{j}_{e}^{mag}$ and $\mathbf{j}_{e}^{con}$ cancel each other out. We have obtained that the total current is independent of $\tau$ in the whole temperature range $T\ll1/\tau$. As a result of this cancellation, the Nernst signal is regular at $T\rightarrow0$. Moreover, the contributions which are constant with respect to the temperature also vanish, and the remaining terms are linear in $T$ in accordance with the third law of thermodynamics.

\section{The phase diagram for the Nernst effect}

\begin{figure}[pt]\vspace{3.5mm}
        \onefigure[width=0.33\textwidth]{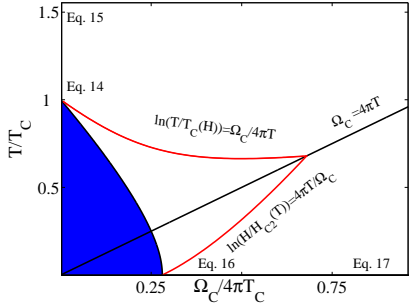}
                 \caption[0.4\textwidth]{\small The phase diagram for the Peltier coefficient $\alpha_{xy}$. We indicate the equations in the text which give the corresponding expressions to $\alpha_{xy}$ in the different limits. $\Omega_c=4eHD/c$ is the cyclotron frequency for the fluctuations of the superconducting order parameter in the diffusive regime.} \label{fig:PhaseDiagram}
\end{figure}

In the following section we present the theoretical expressions for the transverse Peltier coefficient for a superconducting film in the normal state for various regions of the temperature and the magnetic field.  The phase diagram for the Peltier coefficient is plotted in Fig.~\ref{fig:PhaseDiagram}. In the area below the  line  $\ln(T/T_c(H))=\Omega_c/4\pi{T}$ the Landau level quantization of the superconducting fluctuations becomes essential. The line $\ln(H/H_{c_2}(T))=4\pi{T}/\Omega_c$ separates the regions of classical and quantum fluctuations.

For a small magnetic field, $\Omega_c\ll{T}$, close to the transition temperature ($T\approx{T_c}$) the leading contribution to $\alpha_{xy}$ is given by the Aslamasov-Larkin term (see Fig.~\ref{fig:diagrams}(c)) and the magnetization current:
\begin{equation}\label{eq:closetoTc}
\alpha_{xy}\approx\frac{e\Omega_c}{192T\ln{T/T_c(H)}}.
\end{equation}
Note that unlike the electric conductivity, $\sigma _{xx}$, for which the
anomalous Maki-Thompson~\cite{Maki1968} and the Aslamazov-Larkin terms yield
comparable corrections, the contribution from the anomalous Maki-Thompson
term to the Nernst signal is $\sim ({T}/\varepsilon _{F})^{2}\ll 1$ smaller
than the one given by Eq.~\ref{eq:closetoTc}. Therefore, it is natural that
in the vicinity of $T_{c}$ our result coincides with the expression~\cite%
{Ussishkin2002,Varlamov} obtained phenomenologically from the time dependent
Ginzburg-Landau equation (TDGL).

\begin{figure}[t]\center
        \includegraphics[width=0.38\textwidth]{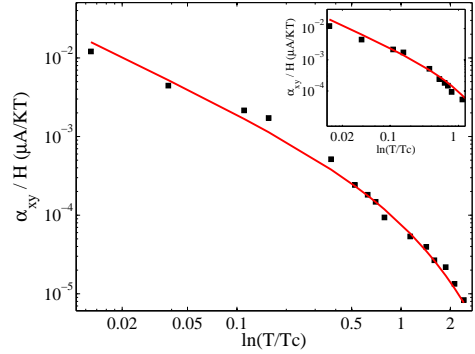}
                 \caption[0.4\textwidth]{\small The transverse Peltier coefficient $\alpha_{xy}$
divided by the magnetic field $H$ as a function of $\ln{T/T_c}$ for a vanishingly small magnetic field. The experimental data of Ref.~\cite{Aubin2007} are presented by the black squares and the solid line corresponds to the theoretical curve given by Eq.~\ref{eq:farfromTc}.  The inset presents the fitting of the data in the vicinity of $T_c$ with Eq.~\ref{eq:closetoTc}.} \label{fig:fittingT=Tc}
\end{figure}

When temperature is increased further away from the critical temperature, the sum of the contributions to the transverse  Peltier coefficient from all the diagrams and the magnetization current yields:
\begin{equation}\label{eq:farfromTc}
\alpha_{xy}\approx\frac{e\Omega_c}{24\pi^2T\ln{T/T_c}}.
\end{equation}
The comparison of our result with the experimental observation of Ref.~\cite{Aubin2007} for a $Nb_{0.15}Si_{0.85}$ film of thickness $35{\rm nm}$ and $T_c=380{\rm mK}$  is given in Fig.~\ref{fig:fittingT=Tc}. The Peltier coefficient  depends on the mean field temperature of the superconducting transition, $T_c^{MF}$, and on the diffusion coefficient through  $\Omega_c$. Throughout the paper we fit the data using the same diffusion coefficient $D=0.187{\rm cm^2/sec}$ which is within the measurement accuracy of the value that has been extracted from the experiment in Ref.~\cite{Aubin2007}.  We take $T_c^{MF}=385{\rm mK}$ which is slightly higher than the measured critical temperature anticipating a small suppression of the temperature of the transition by fluctuations. (An equally well fitting of the data has been obtained for a thinner film of thickness $12.5{\rm nm}$.)

In the vicinity of $T_c$, one can interpret the expression in Eq.~\ref{eq:closetoTc} in terms of the classical picture in which the Cooper pairs with a finite lifetime are responsible for the thermoelectric current. Far from the critical temperature, the quantum nature of the fluctuations reveals itself in contributions to $\mathbf{j}_{e}^{con}$ and $\mathbf{j}_{e}^{mag}$ that are of the order $\ln(\ln1/T\tau)-\ln(\ln{T}/T_c)$. However, these $\tau$-dependent terms in $\mathbf{j}_{e}^{con}$ and in the magnetization current cancel each other out. Thus, the third law of thermodynamics constrains the magnitude of the Nernst signal not only at $T\rightarrow0$ but also at high temperatures, $T\gg{T}_c$.

%

Finally, we present the expressions for the Nernst signal in the high magnetic field part of the phase diagram, $\Omega_c\gg{T}$.  As has been discussed above, after the cancelation of the diverging terms, the remaining contributions to $\alpha_{xy}$ in the limit $T\rightarrow0$ are linear in the temperature:
\begin{equation}\label{eq:alphaB=Bc}
\alpha_{xy}\approx-\frac{eT\ln3}{3\Omega_c(\ln{H/H_{c_2}(T)})^2}\hspace{5mm}\hbox{for $H\approx{H_{c_2}}$},
\end{equation}
and
\begin{equation}\label{eq:alphaB<Bc}
\alpha_{xy}\approx\frac{2eT}{3\Omega_c\ln{H/H_{c_2}}}\hspace{5mm}\hbox{for $H\gg{H_{c_2}}$}.
\end{equation}
Notice that $\alpha_{xy}$  changes its sign in this region. Since the transverse signal is non-dissipative the sign of the effect is not fixed. Although we are in the low temperature limit, as we have already explained, the integrals determining  $\alpha_{xy}$ accumulate at low frequencies. This situation is rather peculiar; it is not typical for fluctuations induced by a quantum phase transition.

\begin{figure}[pt]
        \onefigure[width=0.35\textwidth]{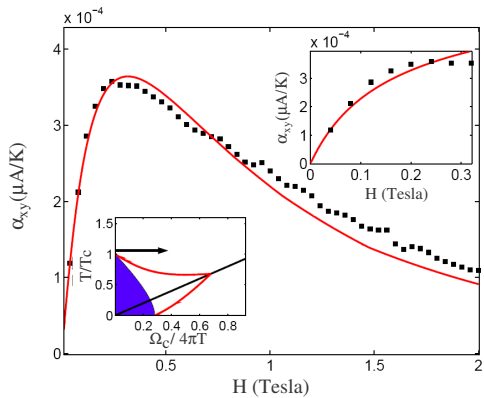}
                 \caption[0.4\textwidth]{\small The transverse Peltier coefficient $\alpha_{xy}$
 as a function of the magnetic field measured at $T=410{\rm mK}$. The black squares correspond to the experimental data of Ref.~\cite{Aubin2007} while the solid line describes the theoretical result. The arrow on the phase diagram illustrates the direction of the measurement. In the inset the low magnetic field data are fitted with the theoretical curve given by Eq.~\ref{eq:closetoTc}.} \label{fig:fittingB}
\end{figure}

In Fig.~\ref{fig:fittingB} we plot the curve for the Peltier coefficient as a function of the magnetic field at a temperature higher than $T_c$. Fig.~\ref{fig:fittingB} demonstrates the agreement between the theoretical expressions and the experimental observation for a broad range of magnetic fields. In addition, we show that the experimental data are well described by Eq.~\ref{eq:closetoTc} in the limit of vanishing magnetic field (see inset of Fig.~\ref{fig:fittingB}). Since Eq.~\ref{eq:closetoTc} is valid in the limit $\Omega_c\ll{T}$,  it can describe only the first few points in the measurement. In order to fit the entire range of the magnetic field, we had to include higher order terms of $\Omega_c/T$. For that, we needed to sum the contributions from all diagrams and the magnetization current. We performed the calculation assuming that  $\ln(T/T_c(H))\ll1$; therefore the theoretical curve starts to deviate from the measured data when $\ln(T/T_c(H))$ is no longer small ($H\approx1Tesla$).

\section{Summary}
In this letter, we have demonstrated that the  fluctuations of the superconducting order parameter are responsible for the large Nernst effect observed in disordered films far away from the transition. Under the condition of the experiment~\cite{Aubin2006,Aubin2007}, the signal generated by the fluctuations dominates the one produced by the quasi-particles up to $T\lesssim100T_c$ and  $H\lesssim100H_{c_2}$. We have outlined the main steps in the derivation of the Nernst effect using the quantum kinetic equation. In this method, one gets automatically all contributions to the Nernst coefficient as response to the temperature gradient, in particular, the one from the magnetization current. We have shown that the important role of the magnetization current is in canceling the quantum contributions, thus making the Nernst signal compatible with the third law of thermodynamics. The third law of thermodynamics constrains the magnitude of the Nernst signal not only at low temperatures, but also far from $T_c$. As a consequence of this constraint the phase diagram is less rich and diverse than one expects in the vicinity of a quantum phase transition.

The Nernst effect provides an excellent opportunity to test the use of the quantum kinetic equation in the description of thermoelectric transport phenomena. We should remark that our results are different in few aspects from the expressions for the Peltier coefficient recently obtained using the Kubo formula in Ref.~\cite{Skvortsov2008}. The striking agreement between our results and the experiment in the different limits (and the fact that we have
reproduced the phenomenological result of the TDGL~\cite{Ussishkin2002}) reinforces us in the correctness of our method.

\acknowledgments
We thank H.~Aubin, K.~Behnia, J.~Sinova, M.~A.~Skvortsov and A.~A.~Varlamov for valuable discussions.
This research was supported by the US-Israel BSF and the Minerva Foundation.

\end{document}